\documentclass[prl,showpacs,footinbib,twocolumn,final]{revtex4}
\usepackage{graphics,graphicx,color}
\begin{document}

\title{Dimensional crossover in quantum networks: from macroscopic to mesoscopic Physics}
\author{F\'elicien Schopfer$^{1,*}$, Fran\c{c}ois Mallet$^{1,2}$, Dominique Mailly$^{3}$, Christophe Texier$^{4,5}$, Gilles Montambaux$^{5}$, Christopher B\"{a}uerle$^{1,2}$, and Laurent Saminadayar$^{1,2,5}$}
\affiliation{$^1$Centre de Recherches sur les Tr\`es Basses Temp\'eratures, B.P. 166 X, 38042 Grenoble Cedex 09, France}
\affiliation{$^{2}$Institut N\'{e}el, B.P. 166 X, 38042 Grenoble Cedex 09, France}
\affiliation{$^3$Laboratoire de Photonique et Nanostructures, route de Nozay, 91460 Marcoussis, France}
\affiliation{$^4$Laboratoire de Physique Th\'{e}orique et Mod\`{e}les Statistiques, Universit\'{e} de Paris XI, 91405 Orsay Cedex, France}
\affiliation{$^5$Laboratoire de Physique des Solides, Universit\'{e}  de Paris XI, 91405 Orsay Cedex, France}
\affiliation{$^6$Universit\'{e} Joseph Fourier, B.P. 53, 38041 Grenoble Cedex 09, France}
\date{\today}

\begin{abstract}
We report on  magnetoconductance measurements  of metallic networks of various sizes ranging from $10$ to $10^{6}$ plaquettes, with anisotropic aspect ratio. Both Altshuler-Aronov-Spivak (AAS) $h/2e$ periodic oscillations and Aharonov-Bohm (AB) $h/e$ periodic oscillations  are observed for all networks. For large samples, the amplitude of both oscillations results from  the incoherent superposition of contributions of phase coherent regions. When the transverse size becomes smaller than the phase coherent length $L_\phi$, one enters a new regime which is phase coherent (mesoscopic) along one direction and macroscopic along the other, leading to a new size dependence of the quantum oscillations.
\end{abstract}
\maketitle

Quantum interference effects lie at the heart of mesoscopic physics. It is well known that they govern both thermodynamic as well as electronic transport properties of quantum conductors. One of the most spectacular manifestations of such quantum interferences is the Aharonov-Bohm effect~\cite{Aharonov59}  in a mesoscopic ring whose perimeter is of the order of the phase coherence length $L_\phi$: when applying a magnetic flux through the ring, the conductance oscillates with a periodicity $\phi_{0}=h/e$, the flux quantum, $h$ being the Planck constant and $e$ the charge of the electron~\cite{Buttiker81}. Such a magnetoconductance oscillation is a direct consequence of the coupling of the electron charge to the vector potential, and is thus the most direct evidence of the quantum nature of the conduction in mesoscopic systems~\cite{Webb86}.

An important point is the understanding of how such quantum effects disappear when going from mesoscopic to macroscopic conductors. If one considers lines of $N$ mesoscopic metallic rings, the Aharonov-Bohm (AB) conductance oscillations $\delta G_{\rm AB}/G$ vanishes to zero as 1/$\sqrt{N}$. This has been beautifully demonstrated by studying lines of silver rings with $N$ varying from $1$ to $30$~\cite{Umbach86}.

On the other hand, there exist magnetoconductance oscillations which \emph{do} survive such an ensemble averaging, since they are due to interferences between time reversed trajectories. These oscillations $\Delta G_{\rm AAS}$, known as Altshuler-Aronov-Spivak (AAS) oscillations, have a period $\phi_{0}/2$~\cite{Altshuler81,Sharvin81}. The robustness of these oscillations towards ensemble averaging, as opposed to the AB oscillations was also experimentally demonstrated in ref.~\cite{Umbach86}.  The relative amplitude $\Delta G_{\rm AAS} / G $ was found to be independent of $N$.  This robustness has also been demonstrated in large two-dimensional metallic networks of different topologies~\cite{Pannetier84,Dolan86}. It must be stressed  that all these experiments have been carried out in a regime where the phase coherence length $L_\phi$ is much smaller than the system size. In this context, one deals with the simple case of an ensemble averaging consisting in a summation of \emph{uncorrelated} contributions from phase coherent regions.

A crucial question is to know what happens to the ensemble averaging when the system size decreases and becomes of the order or smaller than the phase coherence length $L_\phi$. In this Letter, we report on the size dependence of the amplitudes of both Aharonov-Bohm and Altshuler-Aronov-Spivak magnetoconductance oscillations in silver networks of anisotropic aspect ratio. We show that the amplitude of both AB and AAS oscillations exhibit an unexpected dependence with $N$ when the smallest dimension of the network becomes smaller than the phase coherence length: in this case, the network can be considered as a fully coherent object (mesoscopic) in one direction, whereas macroscopic in the other.
\bigskip

\begin{figure}[h]
\includegraphics[width=8.5cm]{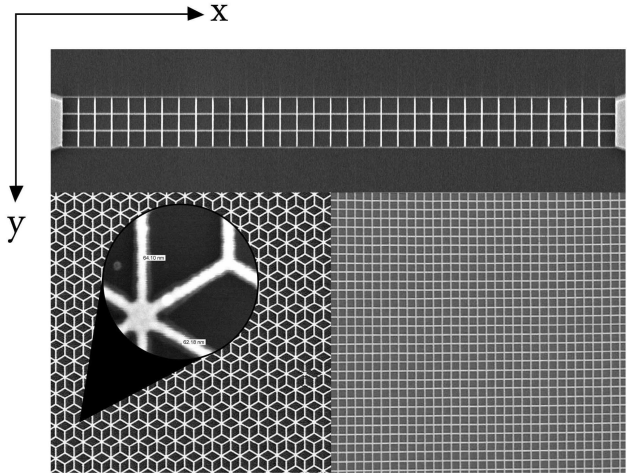}
\caption{Scanning Electron Micrograph of several samples of various sizes; the two  contacts are visible for the small sample.}
\label{Sample}
\end{figure}

Samples are fabricated on a silicon substrate using electron beam lithography on polymethyl-methacrylate resist. Silver is deposited from a $99.9999\,\%$ purity source using an electron gun evaporator and lift-off technique without any additional adherence layer. All samples have been evaporated in a single run to ensure that the sample characteristics (elastic mean free path $l_e$ and phase coherence length $L_\phi$) are similar. Two different topologies have been studied in this work: the square lattice and the so-called $\mathcal{T}_{3}$ lattice~\cite{Naud01}. The wires forming the networks are $60\,nm$ wide, $50\,nm$ thick and $640\,nm$ ($690\,nm$) long for the square (${\mathcal T}_3$) lattice. The size of the plaquettes (square or diamond) is chosen such that the magnetic field corresponding to one flux quantum $\phi_{0}$ per plaquette is $B=100\,G$. All networks with number of plaquettes $N$ varying from $10$ to $10^6$ have the same aspect ratio $L_x/L_y=10$ (see figure~\ref{Sample}). As a consequence, their resistances are similar and of the order of $100\,\Omega$. Measurements have been performed at $400\,mK$; this allows to stay in the linear regime with a relatively high current ($\approx 4\,nA$) and optimizes the signal to noise ratio without heating the electrons. At this temperature, the phase coherence length, determined from standard weak localization measurements on a $120\,nm$ wide wire fabricated on the same wafer, is about $L_\phi\simeq6\,\mu m$, the diffusion constant $D\simeq105\, cm^{2}s^{-1}$ and the thermal length $L_T=\sqrt{\hbar D/k_{B}T}\simeq0.45\,\mu m$~\cite{Mallet05}.

\begin{figure}[t]
\includegraphics[width=9cm]{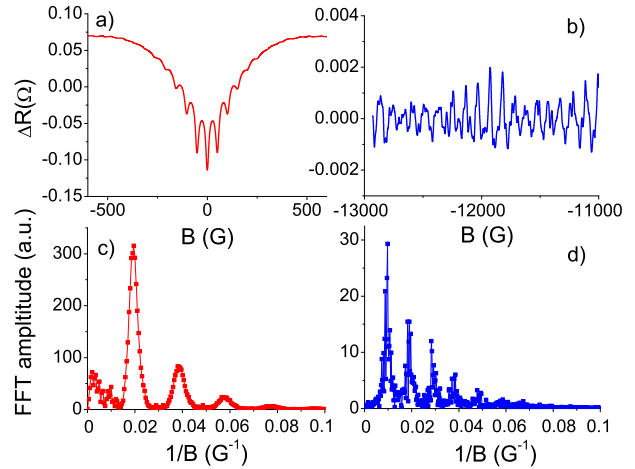}
\caption{(color online). Magnetoresistance of a square network containing 3000 plaquettes: a) low field data; b) high field data; c), d) Fourier amplitudes of a), b) respectively.}
\label{Magneto}
\end{figure}

In figure~\ref{Magneto} we show typical data for the magnetoresistance of a square network with $3000$ plaquettes. At low field (figure~\ref{Magneto}a),  oscillations with a period $B = 50\,G$, corresponding to $\phi_{0}/2$ per plaquette are identified as the AAS oscillations. At fields typically higher than the field which suppresses weak localization, we observe a different type of oscillations. These oscillations have a periodicity of $B = 100\,G$, corresponding to $\phi_{0}$;  these are AB oscillations. In order to emphasize the different periodicity of these magnetoconductance oscillations, we display their Fourier spectra in figures~\ref{Magneto}c (low field) and~\ref{Magneto}d (high field): in the high field regime, the main peak clearly appears at $0.01\,G^{-1}$, whereas in the low field regime, it appears at $0.02\,G^{-1}$. To our knowledge, this is the first time that \emph{both} AAS and AB oscillations are observed on \emph{such large samples}.

We now concentrate on the variation of the amplitude of the AB as well as AAS oscillations \textit{versus} the number of plaquettes $N$. To measure the AB oscillations we sweep the magnetic field from $7000\,G$ to $13000\,G$, whereas for the AAS oscillations we cover a field range of $\pm 1200\,G$. To extract precisely the amplitude of the AB oscillations, we take the Fourier transform over $20$ periods   after subtraction of a smooth background to remove low frequency fluctuations. We also measure the background noise  by repeating the measurement \emph{exactly in the same conditions} but at fixed magnetic field, and taking again the Fourier transform. The amplitude of the AB signal is then obtained from the Fourier spectrum after subtraction of the background spectrum integrated over the same frequency range, in a similar way used for persistent current measurements~\cite{Rabaud01}. This procedure is only necessary for very large networks (typically larger than $10^{5}$ plaquettes) since for smaller networks the noise is negligible. For the determination of the AAS amplitude such a procedure is not necessary, as the background noise is always negligible. However, the second harmonic ($\phi_0/2$) of the AB oscillations has the same frequency as the first harmonic ($\phi_0/2$) of the AAS oscillations. For small networks (typically $N \leq 100$) this contribution cannot be neglected. In order to extract the AAS signal, we therefore determine first the amplitude of the second harmonic of the AB oscillations at high field and then subtract this amplitude from the first harmonic of the oscillations measured at low field \cite{Phase}.

\begin{figure}[t!]
\includegraphics[width=9cm]{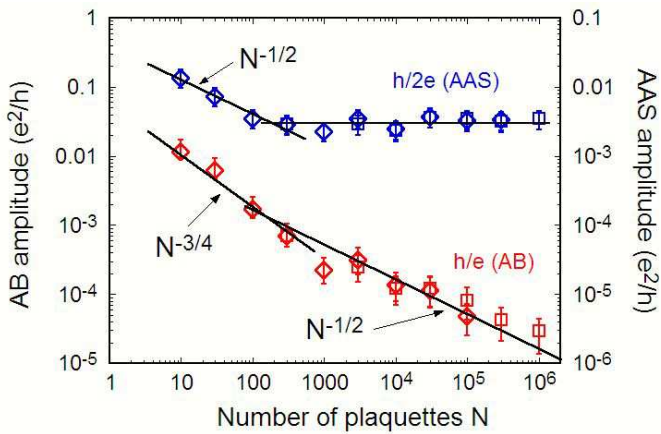}
\caption{(color online). AAS amplitude $\Delta g_{\rm AAS}$ and AB amplitude $\delta g_{\rm AB}$ as a function of the number  $N$ of plaquettes for different networks of various sizes $L_x\times L_y\propto N$, for two topologies, square ({\it squares}) and ${\cal{T}}_3$ ({\it diamonds})~\cite{Amplitude}.}
\label{DeltaG1}
\end{figure}

In figure~\ref{DeltaG1} we display the amplitude  of magnetoconductance oscillations (AAS and AB) extracted from the Fourier spectra as a function of the number  $N$ of plaquettes. For large networks ($N\gtrsim 300$), the amplitude of the AB oscillations clearly decreases as $1/\sqrt{N}$, whereas the amplitude of the AAS oscillations are independent on the number of plaquettes as naively expected.
More surprising is the behavior observed for small networks: when they contain typically less than $N \simeq 300$ plaquettes, the amplitude of the AB oscillations  varies  faster than $1/\sqrt{N}$. At the same time the AAS amplitude now depends on $N$ (figure~\ref{DeltaG1}). In the following, we will show that this new behavior results from a dimensional crossover when the transverse size of the network becomes smaller than  the phase coherence length: one then enters a new regime where the transport properties are effectively one-dimensional on the two-dimensional network.

Let us first recall general considerations on the magnetoconductivity oscillations: on one hand, the AAS oscillations are the Fourier harmonics of the weak localization correction $\langle \Delta \sigma(B) \rangle$ to the average conductivity~\cite{Altshuler81}. On the other hand, the amplitude of the AB oscillations can be obtained from the correlation function of the conductivity $\langle\delta\sigma(B)\delta\sigma(B')\rangle$. The expressions of these two quantities are indeed related~\cite{AleBla02,LM,TexMon04,Montambaux04}. In the limit where the thermal length $L_T$ is smaller than $L_\phi$, this yields
\begin{equation}
\label{exp1}
\delta\sigma_{\rm AB}^2={e^2 \over h}\, {4 \pi L_T^2 \over 3 \,\rm{Vol}}\, \Delta\sigma_{\rm AAS}
\end{equation}
where $\delta \sigma_{\rm AB}$  and $\Delta \sigma_{\rm AAS}$ are respectively the first harmonics of the AB and AAS oscillations and Vol the volume of the sample.   In this formula, the  temperature dependence originates from the  conductivity correlation function which probes a finite energy scale of width $k_B T$~\cite{Webb86,Montambaux04}. The key feature is the proportionality between $\delta \sigma_{\rm AB}^2$ and $\Delta \sigma_{\rm AAS}$. Indeed, both quantities can be written in terms of the coherent part of the return probability to the origin for a diffusive particle. Consequently both must probe in the same way the influence of the geometry~\cite{Montambaux04}.

We consider a network of dimensions $L_x \times L_y$ (see figure~\ref{Sample}). It is important to keep in mind that experiments presented here are performed on several networks of different sizes, but \emph{of constant aspect ratio} $L_x/L_y=10$. The length and width of the networks thus scale with the number of plaquettes $N$ as $L_x \propto L_y \propto\sqrt{N}$.

The dimensionless conductance $g=G/(2 e^2 /h)$ of the network is then related to the conductivity by Ohm's law $g\propto\sigma L_{y} /L_{x}  $. Combined with eq.~(\ref{exp1}) and  given that $\mbox{Vol} \propto L_x \, L_y$, this yields for the amplitudes of the conductance oscillations $\Delta g_{\rm AAS}$ and $\delta g_{\rm AB}$:
\begin{equation}
\label{rela1}
\delta g_{\rm AB}^2 = \frac{2\pi L_T^2}{3 L_x^2} \Delta g_{\rm AAS}
\end{equation}
which is the key relation from which we now discuss our results, bearing in mind that   temperature, and thus $L_T$ and $L_\phi$ are fixed parameters.

Let us first consider large networks with both dimensions larger than the phase coherence length:  $ L_x,\,L_y  \gg L_\phi$. {\it Since interfering time reversed trajectories extend over a typical size $L_\phi$, they do not feel the boundaries of the system and therefore $\Delta\sigma$ is size independent.} Therefore the AAS amplitude varies as $\Delta g_{\rm AAS}\propto L_y/L_x$ and since this ratio is constant, this amplitude is independent on $N$:
\begin{equation}
\Delta g_{\rm AAS}\propto N^0
\end{equation}
In this regime we also see from eq.~(\ref{rela1}) that $\delta g_{\rm AB}^2\propto L_y/L_x^3$, which leads to:
\begin{equation}
\delta g_{\rm AB}\propto N^{-1/2}
\end{equation}
This is exactly what is observed for large networks: when the number of plaquettes is larger than $\simeq 300$, electrons diffuse on what they feel as a two-dimensional network.

For smaller networks, the transverse dimension $L_y$ eventually becomes smaller than the phase coherence length: we enter a regime where the network becomes transversally coherent whereas it remains longitudinally incoherent: $L_y\ll L_\phi\ll L_x$. In this case, we have the usual quasi-1D scaling $\Delta \sigma_{\rm AAS} \propto L_\phi/L_y$. Therefore we find $\Delta g_{\rm AAS}\propto 1/L_x$ and $\delta g_{\rm AB}^2\propto 1/L_x^3$, which leads to:
\begin{eqnarray}
\label{scalingmeso}
\Delta g_{\rm AAS}\propto N^{-1/2}\\
\label{scalingmeso2} \delta g_{\rm AB}\propto N^{-3/4}
\end{eqnarray}
This is precisely what is observed for small networks on  figure~\ref{DeltaG1}.

It remains now to check whether the position of the crossover observed on figure~\ref{DeltaG1} agrees with our estimate of the phase coherence length. The crossover occurs for a size $N \simeq 300$ corresponding  to $L_y \simeq 3.8 \,\mu m$. This length has to be compared with   the  coherence length $L_\phi \simeq 6\,\mu m$ measured at $T=400\, mK$. This comparison, which cannot be more than qualitative, supports our analysis.

To summarize, the dimensional crossover observed for the scaling of the AB oscillations corresponds to the different scaling $L_y /L_x^3 \rightarrow L_\phi/L_x^3$ of the variance of the conductance fluctuations, with $N \propto L_x L_y$. At this point, it is useful to compare these dependences with the case of a $1D$ chain where the number $N$ of rings scales linearly with the length $L_x$ of the chain, so that $\delta g_{\rm AB}^2 \propto 1/L_x^3 \propto 1/N^{3}$ and $\Delta g_{\rm AAS} \propto 1/N$. Since the conductance $g$ scales as $1/N$, this yields for the relative fluctuations $\Delta g_{\rm AAS}/g\propto N^0$ and $\Delta g_{\rm AB}/g \propto 1/\sqrt{N}$ as was observed experimentally~\cite{Umbach86}.

\begin{figure}[t]
\includegraphics[width=8.5cm]{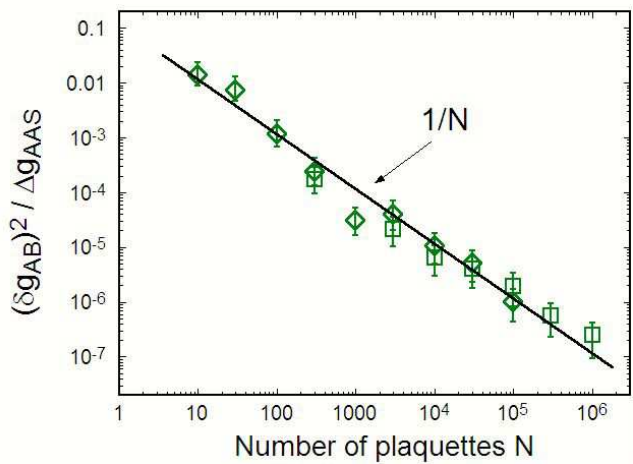}
\caption{(color online). $\Delta g_{\rm AAS}/({\delta g}_{\rm AB})^2$ as a function of the number of plaquettes $N$ for square networks (\textit{squares}) and diamonds networks (\textit{diamonds}).}
 \label{DeltaG2}
\end{figure}

An interesting way of checking our analysis comes from equation~(\ref{rela1}): we can see that the ratio $\delta g_{\rm AB}^2 / \Delta g_{\rm AAS}  \propto L_T^2/L_x^2$ is proportional to $1/N$ and more importantly is independent on $L_\phi$. This fundamental relation between $\delta g_{\rm AB}^2$ and $\Delta g_{\rm AAS}$ is clearly shown on figure~\ref{DeltaG2}, where we have plotted the ratio $\delta g_{\rm AB}^2 / \Delta g_{\rm AAS} $ as a function of the number of plaquettes $N$: one sees that it follows perfectly the predicted $1/N$ behavior, with no dimensional crossover. This is a definitive check of our interpretation of the experimental data in terms of dimensional crossover.
\bigskip

In conclusion, we have measured both Aharonov-Bohm $\phi_{0}$ periodic oscillations and Altshuler-Aronov-Spivak $\phi_{0}/2$ periodic oscillations in metallic networks containing $10$ to $10^{6}$ plaquettes.  Ensemble averaging can lead to different size dependences for small and large networks. The crossover takes place when the width of the network is of the order of the phase coherence length; this behavior does correspond to a dimensional crossover between effectively one- and two-dimensional networks. In this new one-dimensional regime we observed, we have shown that the amplitude of the AB oscillations varies as $N^{-3/4}$ and the AAS oscillations as $N^{-1/2}$, a behavior which has never been observed up to now. Moreover, we have been able to probe experimentally the fundamental relation between AB and AAS magnetoconductance oscillations due to their common physical origin.
\bigskip

\acknowledgments We are indebted to the Quantronics group  for the use of its evaporator and silver source. It is our pleasure to acknowledge  H.~Bouchiat, B.~Dou\c{c}ot,  L.~P.~L\'{e}vy and J.~Vidal for fruitful discussions. This work has been supported by the French Ministry of Science, grants \# 02 2 0222 and \# NN/02 2 0112, and the European Commission FP6 NMP-3 project 505457-1 \textquotedblleft Ultra-1D\textquotedblright.


\begin{thebibliography}{99}
\bibitem{Phase} The relative phase ($0$ or $\pi$) between the AAS oscillations and the second harmonic of the AB oscillations, however, is determined at low field (typically at the field which suppresses weak localization) such that any phase shift due to the applied magnetic field is negligible.
\end{thebibliography}
\end{document}